\documentclass[aps,twocolumn,groupedaddress,showpacs]{revtex4-1}
\usepackage{graphics}
\usepackage{epstopdf}
\usepackage{amsmath,amsfonts,amsthm}
\usepackage{newpxtext,newpxmath}
\usepackage[final]{graphicx}
\usepackage{xcolor}
\usepackage{soul}
\begin{document}
\newcommand{\op}{\boldsymbol}

\title{Wave Particle Duality in Asymmetric Beam Interference}
\thanks{Published: \large\bf Phys. Rev. A 98, 022130 (2018) }

\author{Keerthy K. Menon}
\email{keerthykm1@gmail.com}
\affiliation{Department of Physics,
Central University of Karnataka, Karnataka, India.}
\author{Tabish Qureshi}
\email{tabish@ctp-jamia.res.in}
\affiliation{ Centre for Theoretical Physics,
Jamia Millia Islamia,
New Delhi - 110025.}

\begin{abstract}
It is well known that in a two-slit interference experiment, acquiring
which-path information about the particle, leads to a degrading of the
interference. It is argued that path-information has a meaning only
when one can umabiguously tell which slit the particle went through.
Using this idea, two duality relations are derived for the general case where
the two paths may not be equally probable, and the two slits may be of
unequal widths. These duality relations, which
are inequalities in general, saturate for all pure states. Earlier known
results are recovered in suitable limit.
\end{abstract}

\maketitle

\section{Introduction}

Wave-particle duality is an intriguing aspect of nature, which was first
conceptualized by Neils Bohr in his principle of complementarity \cite{bohr}.
It has been in debate right from the beginning when Einstein's raised
objections against it, proposing his famous recoiling slit experiment
\cite{tqeinstein}, and continues to be so even today \cite{liu}. The two-slit
interference experiment has become a cornerstone for investigating such issues.

Bohr had emphasized that the wave and particle natures are mutually exclusive,
revealing one, completely hides the other.
Wooters and Zurek began by asking what happens if one probes the wave and
particles natures at the same time, in a two-slit experiment \cite{wootters}.
They found that it is 
indeed possible to partially reveal both the natures. This idea was later
put on firm mathematical ground by Englert, in the form of a duality
relation which puts a bound on how much of each nature can be revealed 
simultaneously: \cite{englert}
\begin{equation}
{\mathcal D}^2 + {\mathcal V}^2 \le 1,
\label{englert}
\end{equation}
where ${\mathcal D}$ is path distinguishability, a measure of the particle
nature, and ${\mathcal V}$ the visibility of interference, a measure of
wave nature. Wave and particle natures are so fundamental to
quantum objects that many prefer to call them {\em quantons} \cite{bunge,levy}.
A different kind of duality relations are also studied  where one tries to
{\em predict} the path information of the quanton based on the asymmetry of
the two beams, without using any path detector \cite{greenberger,vaidman}.

Contemporary thinking is that, in a two-slit interference experiment, if
one is able to tell which of the two slits the quanton went through, one
has revealed the particle nature of the quanton. On ther other hand,
if one obtains an interference patters, one has revealed the wave aspect 
of the quanton. The duality relation derived by Englert was for a
symmetric two-slit experiment, in which the quanton is equally likely to
go through both the slits. However, there can be situations where the
setup is not symmetric, i.e., the state of the incident quanton is such that
the probabilities to go through the two slits are not equal. Another
possibility is that the two slits may not be of equal widths. This 
asymmetric case has not been probed in as much detail as the symmetic
case \cite{englert}. There have been other studies on asymmetric
two-path interference \cite{lili,yliu}, but none provides a tight duality
relation for this case. The aim of this paper is obtain a general duality
relation which also holds for asymmetric two-slit interference experiments.

\section{Dealing with asymmetry}

We assume that the state of the quanton that emerges from the double-slit is
given by the {\em unnormalized} state
\begin{equation}
 |\psi\rangle = \sqrt{p_1}|\psi_1\rangle + \sqrt{p_2}|\psi_2\rangle,
\label{asym}
\end{equation}
where $p_1, p_2$ quantify the asymmetry of the incoming wave. In addition,
we would like to take into account the effect of asymmetric slits, namely
the situation where the two slits may have different widths. As different
slit widths will also contribute to the probabilities of passing through
the two slits, one should
assume that the states $|\psi_1\rangle, |\psi_2\rangle$,
corresponding to the quanton coming out of slit 1 and 2, respectively,
are {\em not normalized}. One can see that for the case $p_1=p_2=1/2$,
the probability of the quanton to pass through slit 1 and 2 is proportional
to $\langle\psi_1|\psi_1\rangle$ and $\langle\psi_2|\psi_2\rangle$,
respectively. 

Given that the incoming quanton state is symmetric, the quanton is more likely
to pass through the wider slit. The details of the effect of asymmetry due to
the slits will be specfied while choosing the form of
$|\psi_1\rangle, |\psi_2\rangle$.
Needless to say, the state $|\psi\rangle$ as a whole should be normalized.

\section{Getting which-way information}

An experiment to find out which of the two paths a quanton has followed,
would require some kind of detector which can retrieve and store information
on which path a particular quanton took. We assume a fully quantum
detector with states corresponding to the quanton taking one path or the
other. Without going into the details of what this path detector should
be like, we just use von Neumann's criterion for a quantum measurement
\cite{neumann}. According to von Neumann's criterion, the basic requirement 
for a path detector to perform a which-path measurement is that it should
interact with the quanton in such a way that its two states should get
correlated with the two paths of the quanton. If the state of the quanton
that emerges from the asymmetric double-slit is given by (\ref{asym}),
and the path-detector is in an initial state $|d_0\rangle$, the interaction
between the two should be such that it evolves to the following:
\begin{equation}
 (\sqrt{p_1}|\psi_1\rangle + \sqrt{p_2}|\psi_2\rangle)|d_0\rangle \rightarrow
 \sqrt{p_1}|\psi_1\rangle|d_1\rangle + \sqrt{p_2}|\psi_2\rangle|d_2\rangle
\end{equation}
The quanton goes and registers on the screen, and the path-detector is left
with the experimenter. If the experimenter finds that the state of the
path-detector is $|d_1\rangle$, she can conclude that the quanton went through
slit 1, else if the state of the path-detector is $|d_2\rangle$, it would
imply that the quanton went through slit 2. The interaction between the
quanton and the path-detector is designed by the experimenter, and thus the
states $|d_1\rangle$, $|d_2\rangle$ are known. What is not known is, which
of the two states one would get, for particular instance of quanton
going through the double-slit.

The problem of finding which path the quanton
followed then reduces to telling whether the state of the path-detector
is $|d_1\rangle$ or $|d_2\rangle$. To solve this problem, Englert took the
approach of calculating the optimum ``likelihood for guessing the way
(which of the two ways the quanton went) right". We take a somewhat 
different route. We believe that for any given instance of quanton passing
through the double-slit, one can claim to have which-path knowledge only
if one can tell {\em for sure} which of the two paths the quanton took.
What we mean is, there should be no guessing involved. One should have an
unambigous answer to the question which path the quanton took. In the
path-detection scenaio discussed above, this would mean one should be
able unambiguously tell which of the two states $|d_1\rangle$ or $|d_2\rangle$,
is the given state of the path-detector. If $|d_1\rangle$ and $|d_2\rangle$
are orthogonal, one can measure an observable of the path-detector which
has $|d_1\rangle$ and $|d_2\rangle$ as it's two eigenstates, with distinct
eigenvalues. Looking at the eigenvalue of the measurement, one would know
{\em for sure} that the path-detector was (say) $|d_2\rangle$, and hence
the quanton went through the lower slit. However, there are situations in
which $|d_1\rangle$ and $|d_2\rangle$ are not orthogonal. There exists a
method which allows for unambiguously distinguishing between two
non-orthogonal states, and goes by the name of unambiguous quantum state
discrimination (UQSD) \cite{uqsd,dieks,peres,jaeger2}. A downside of this method is that
there will be occasions where the method will fail to provide an answer,
but the experimenter will know that it has failed. Thus, on the occasions
on which UQSD succeeds, it can unambiguously distinguish between the two 
non-orthogonal states. The measurement method can be tuned to minimize
the failure probability, and thus maximizing the probability of unambiguously
distingishing between the two states.

\section{Unambiguous path discrimination}

The UQSD approach has been successfully used in defining a new
distinguishability of paths, $\cal{D}_Q$, as the maximum probability of 
{\em unambiguously} distinguishing between the available quanton paths.
This resulted in new duality relations for the symmetric two-slit interference
\cite{3slit}, three-slit interference \cite{3slit}, and n-slit interference
\cite{cd,nslit}. Here we use it to study wave-particle duality in the case
of interference involving asymmetric paths.

\begin{figure}[h!]
\centering
\includegraphics[width=8.0 cm]{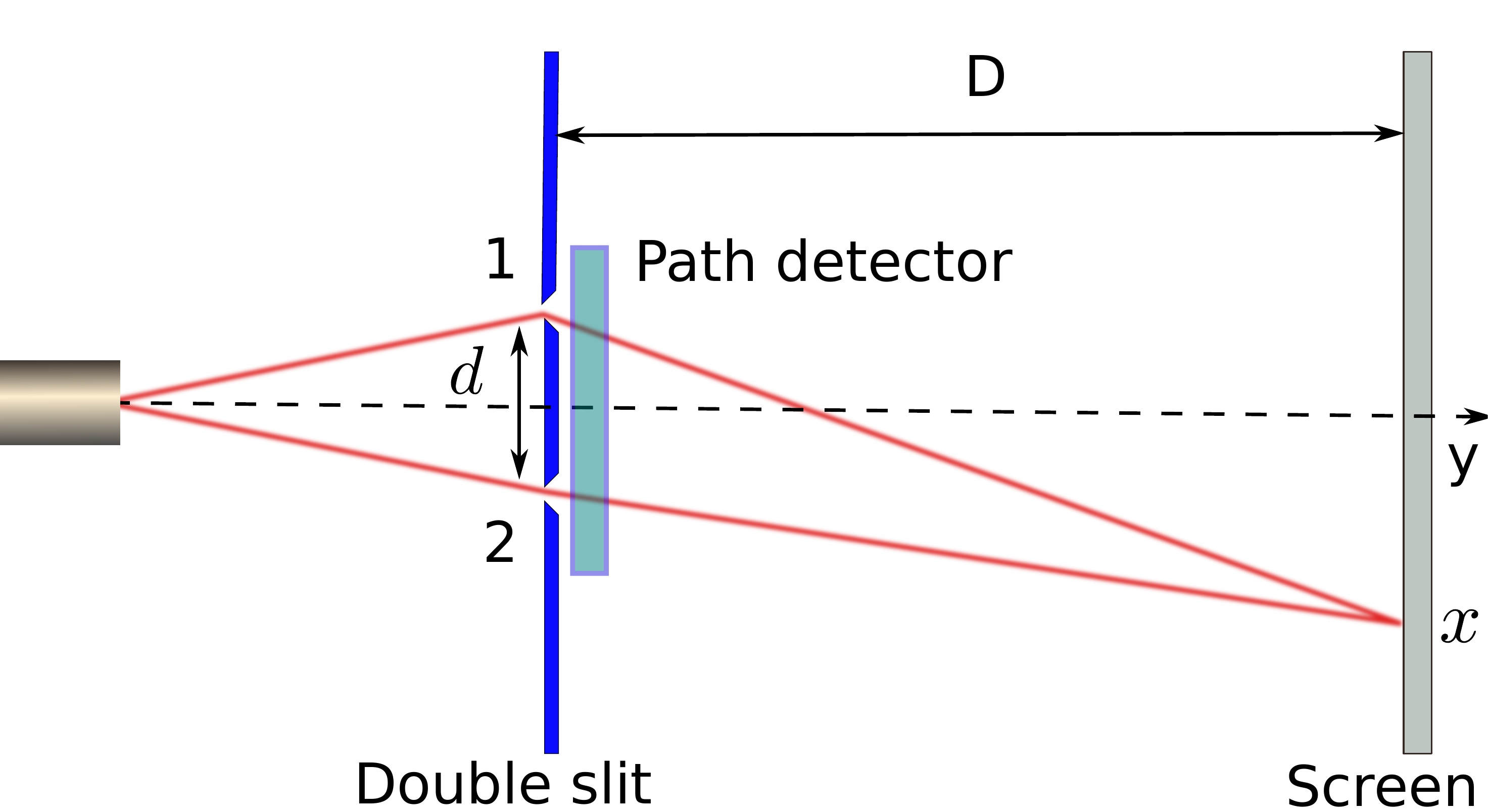}
\caption{A two-slit experiment with a path-detector in front of the double-slit.
Slit separation is $d$ and the distance between the double-slit and the screen
is $D$.}
\label{2slit}
\end{figure}

We begin at the instance the quanton emerges from the double-slit. The
state of the quanton has to be a superposition of two localized parts, in
front of slit 1 and 2, respectively. We assume the quanton is traveling
along the positive y-axis, and the double-slit is the in the x-z plane,
at $y=0$ (see FIG. \ref{2slit}). For the purpose of interference, the motion along the y-axis
is unimportant. It is the spread of the two emerging wave-packets along
the x-axis and the overlap, which gives rise to interference. We neglect the
dynamics along y-axis, and assume that the quanton is traveling along
y-axis with an average momentum $p_0$, and that motion only serves to translate
the quanton from the slit to the screen with time. For calculational
simplicity, we assume the parts of the state emerging from the double-slit
to be Gaussian wave-packets, localized in front of the two slits, namely
at positions $x=x_0$ and $x=-x_0$. The state of the quanton at time $t=0$,
is given by
\begin{equation}
\langle x|\psi(0)\rangle = A\left( \sqrt{p_1}e^{-\frac{(x-x_0)^2}{\epsilon^2}}
+ \sqrt{p_2}e^{-\frac{(x+x_0)^2}{\xi^2\epsilon^2}}\right)
\end{equation}
where $A = \left(\frac{2}{\pi\epsilon^2(p_1+\xi p_2)}\right)^{1/4}$,
$d=2x_0$ is the separation between the slits and $\epsilon$ and $\xi\epsilon$
are the
widths of the two Gaussians, and may loosely be considered the widths of 
the two slits.
At the instant of emerging from the double-slit, the quanton interacts with
a path-detector, and the combined state of the two should have the following
form (as argued earlier):
\begin{equation}
\langle x|\psi(0)\rangle = A \left( \sqrt{p_1}e^{-\frac{(x-x_0)^2}{\epsilon^2}}|d_1\rangle
+ \sqrt{p_2}e^{-\frac{(x+x_0)^2}{\xi^2\epsilon^2}}|d_2\rangle\right),
\label{istate}
\end{equation}
where $|d_1\rangle, |d_2\rangle$ are the two states of the path-detector.
The states $|d_1\rangle, |d_2\rangle$ are chosen to be normalized, although
they are not orthogonal in general. It may be mentioned that choosing 
the probability amplitudes $\sqrt{p_1}, \sqrt{p_2}$ to be real and positive
is not a loss of generality as $|d_1\rangle, |d_2\rangle$ may contain 
phases.

Now, the idea is to find out how much path information of the quanton 
can be retrieved from the path detector, {\em in principle}, given the
state (\ref{istate}). We would like to stress the point that a particular
method of probing the path-detector may yield a certain amount of path
information, but we are interested in best possible value that can be
obtained in principle. UQSD works for the situation where the two
states, $|d_1\rangle, |d_2\rangle$ occur randomly with different 
probability. If one is given one of the states and asked to tell which
of the two it is, UQSD allows one to give the best possible answer.
To use this method for the problem at hand, we should ascertain the
probabilities with which $|d_1\rangle, |d_2\rangle$ occur.
Looking at (\ref{istate}) one may naively jump to the conclusion that the
probabilities in question are $p_1$ and $p_2$. However, the different
widths of the two slits would also contribute to the probability of
the quanton passing through slit $k$ resulting the path-detector state
$|d_k\rangle$. The probability amplitude for this possibility is given by
$c_k = \frac{\langle\psi_k|\psi(0)\rangle}{\sqrt{\langle\psi_k|\psi_k\rangle
\langle\psi(0)|\psi(0)\rangle}}$, where $k=1,2$. Using the Gaussian form
given in (\ref{istate}), these probability amplitudes turn out to be
\begin{eqnarray}
c_1 = {\sqrt{p_1}\over \sqrt{p_1+\xi p_2}} ~~~~
c_2 = {\sqrt{\xi p_2}\over\sqrt{p_1+\xi p_2}}.
\label{c1c2}
\end{eqnarray}
As far as measurements on the path-detector are concerned, it can be assumed
to randomly found in the state $|d_1\rangle$ with probability $c_1^2$, and in
the state $|d_2\rangle$ with probability $c_2^2$. In addition,
without loss of generality, we assume that $c_1 \ge c_2$. 

In order to use UQSD, we assume that the Hilbert
space of the path-detector is not two dimensional, but three dimensional,
described by an orthonormal basis of states $|q_1\rangle, |q_\rangle, |q_3\rangle$.
The reason for doing so will become clear in the following analysis.
The basis is chosen in such a way that the detector states $|d_1\rangle,
|d_2\rangle$ can be represented as \cite{jaeger2}
\begin{eqnarray}
|d_1\rangle &=& \alpha|q_1\rangle + \beta|q_3\rangle \nonumber\\
|d_2\rangle &=& \gamma|q_2\rangle + \delta|q_3\rangle ,
\label{d1d2}
\end{eqnarray}
where $\alpha$ and $\gamma$ are real, and $\beta, \delta$ satisfy
\begin{eqnarray}
|\beta| |\delta| &\ge& |\langle d_1|d_2\rangle|,\nonumber\\
|\beta|^2&=& \max\{|\langle d_1|d_2\rangle|c_2/c_1, |\langle d_1|d_2\rangle|^2\}
\label{betagamma}
\end{eqnarray}
In the expanded Hilbert space, one can now measure an operator (say)
\begin{equation}
\op{A} = |q_1\rangle\langle q_1| + 2|q_2\rangle\langle q_2|
+ 3|q_3\rangle\langle q_3|.
\end{equation}
It is straightforward to see that getting eigenvalue 1 means the state was
$|d_1\rangle$, getting eigenvalue 2 means the state was $|d_2\rangle$.
However, there is also a finite probability of getting eigenvalue 3,
in which case one cannot tell if the state was $|d_1\rangle$ or $|d_2\rangle$.
One would like to minimize the probability of getting the eigenvalue 3,
or failure of the state discrimination. It can be shown that chosen values of
$\beta,\delta$ in (\ref{betagamma}) are such that they minimize the
probability of failure, and maximize the probability of successfully
distinguishing between $|d_1\rangle$ and $|d_2\rangle$ \cite{jaeger2}.
We will return to these in more detail later.

\section{Interference and fringe visibility}

We now analyze what happens when the quanton reaches the screen. We assume
that the quanton takes a time $t$ to travel along y-axis from the double-slit
to the screen, a distance $D$ (see FIG. \ref{2slit}). The time evolution depends on what is the 
nature of our quanton. It could be a photon traveling with the speed of light,
or it could be a particle of mass $m$ under free time-evolution.
One can write the time evolution of the state in a universal form
\begin{equation}
|\psi(t)\rangle = {1\over 2\pi}\int_{-\infty}^{\infty}
|k\rangle \langle k|\psi(0)\rangle e^{-i\omega_kt} dk
\end{equation}
where $|k\rangle$ are the momentum states. For photons $\omega_k = ck$
and for massive particles $\omega_k = \frac{\hbar k^2}{2m}$.
The state of the quanton, after a time $t$ (after traveling a distance $D$
from the double-slit to the screen), can be worked out to be \cite{dillon}
\begin{equation}
\langle x|\psi(t)\rangle = B\left(\tfrac{\sqrt{p_1}}{\sqrt[4]{\epsilon^4+\Gamma^2}}e^{-\frac{(x-x_0)^2}{\epsilon^2+i\Gamma}}|d_1\rangle
+ \tfrac{\sqrt{\xi p_2}}{\sqrt[4]{\xi^4\epsilon^4+\Gamma^2}}e^{-\frac{(x+x_0)^2}{\xi^2\epsilon^2+i\Gamma}}|d_2\rangle\right),
\label{fstate}
\end{equation}
where $\Gamma = 2\hbar t/m = \lambda D/\pi$, if one defines $\lambda = h/p_0$,
and $B = \sqrt[4]{\frac{2\epsilon^2}{\pi(p_1+\xi p_2)}}$.
It can be shown that if the quanton is a photon, one gets the same
expression with $\Gamma = \lambda D/\pi$, where $\lambda$ is the wavelength
of the photon.
Let us assume a phase factor associated with the detector states:
$\langle d_1|d_2\rangle = |\langle d_1|d_2\rangle|e^{i\theta}$.
The probability of the quanton to arrive at a position $x$ on the screen is
then given by
\begin{eqnarray}
|\langle x|\psi(t)|^2 &=& B^2\left(\tfrac{p_1}{\sqrt{\epsilon^4+\Gamma^2}}e^{-\frac{2\epsilon^2(x-x_0)^2}{\epsilon^4+\Gamma^2}}
+ \tfrac{\xi p_2}{\sqrt{\xi^4\epsilon^4+\Gamma^2}} e^{-\frac{2\xi^2\epsilon^2(x+x_0)^2}{\xi^4\epsilon^4+\Gamma^2}}\right.\nonumber\\
&+&\left. \tfrac{\sqrt{\xi p_1p_2}}{\sqrt[4]{\epsilon^4+\Gamma^2}\sqrt[4]{\xi^4\epsilon^4+\Gamma^2}} |\langle d_1|d_2\rangle|
e^{-\frac{(x-x_0)^2}{\epsilon^2-i\Gamma}} 
      e^{-\frac{(x+x_0)^2}{\xi^2\epsilon^2+i\Gamma}+i\theta}\right. \nonumber\\
&+&\left. \tfrac{\sqrt{\xi p_1p_2}}{\sqrt[4]{\epsilon^4+\Gamma^2}\sqrt[4]{\xi^4\epsilon^4+\Gamma^2}} |\langle d_1|d_2\rangle|
e^{-\frac{(x-x_0)^2}{\epsilon^2+i\Gamma}} 
      e^{-\frac{(x+x_0)^2}{\xi^2\epsilon^2-i\Gamma}-i\theta} \right).\nonumber\\
\end{eqnarray}
In the Fraunhofer limit $\lambda D \gg \epsilon^2$, which implies
$\Gamma^2 \gg \epsilon^4$, the above can be simplified to
\begin{eqnarray}
|\langle x|\psi(t)|^2 &=& \tfrac{B^2}{\Gamma} \left[p_1e^{-\frac{2\epsilon^2(x-x_0)^2}{\Gamma^2}}
+ \xi p_2 e^{-\frac{2\epsilon^2(x+x_0)^2}{\Gamma^2}} \right.\nonumber\\
&&\left. +~ 2\sqrt{p_1p_2\xi} |\langle d_1|d_2\rangle| 
e^{-\frac{\epsilon^2(x^2+x_0^2)(1+\xi^2)}{\Gamma^2}}
e^{\frac{\epsilon^2 2xx_0(1-\xi^2)}{\Gamma^2}}
\right. \nonumber\\
&&\times\left. \cos\left({\frac{4\pi xx_0}{\lambda D}+\theta}\right)\right].
\label{interf}
\end{eqnarray}
Eqn. (\ref{interf}) represents a two-slit interference pattern, with a 
fringe width $w = \lambda D/d$. We assume that tha intensity at position $x$
is given by $I(x) \propto |\langle x|\psi(t)|^2$. The maxima and minima of
intensity occur at the values of $x$ where the cosine term is 1 and -1,
respectively.

The visibility of the interference pattern is just the 
the contrast in intensities of neighbouring fringes \cite{born}
\begin{equation}
{\mathcal V} = \frac{I_{\rm{max}} - I_{\rm{min}}}{ I_{\rm{max}} + I_{\rm{min}} } ,
\end{equation}
where $I_{\rm{max}}$ and $I_{\rm{min}}$ represent the maximum and minimum
intensity in neighbouring fringes. The interference in (\ref{interf}),
ignoring the effect of a finite slit-width $\epsilon$, yields
the following {\em ideal} fringe visibility:
\begin{equation}
{\mathcal V} = \frac{2\sqrt{p_1p_2\xi}}{p_1+\xi p_2} |\langle d_1|d_2\rangle|.
\label{V}
\end{equation}
If $|d_1\rangle=|d_2\rangle$, which means that the path-detector is
effectly absent, the fringe visibility reduces to
${\mathcal V}_0 = \frac{2\sqrt{p_1p_2\xi}}{p_1+\xi p_2}$, and is called the {\em a priori} fringe
visibility. This means that even if no which-path information is extracted,
the visibility will be less than 1 if either the incoming state is asymmetric,
or the slits are of unequal widths.

\section{Distinguishability \& duality relations}

Coming back to the issue of getting unambiguous path information about the
quanton, notice that (\ref{betagamma}) implies two cases:
(a) $|\beta|^2 = |\langle d_1|d_2\rangle|c_2/c_1 \ge |\langle d_1|d_2\rangle|^2$
and
(b) $|\beta|^2 = |\langle d_1|d_2\rangle|^2 > |\langle d_1|d_2\rangle|c_2/c_1$.
These should be discussed separately.
We define the distinguishability of two paths, ${\mathcal D}_Q$, as the
maximum probability with which the two paths can be {\em unambiguously}
distinguished. To get distinguishability, we first use (\ref{d1d2}) to
rewrite (\ref{fstate}) as
\begin{eqnarray}
|\psi(t)\rangle &=& c_1|\psi_1(t)\rangle|d_1\rangle
+ c_2|\psi_2(t)\rangle|d_2\rangle \nonumber\\
&=& c_1\alpha|\psi_1(t)\rangle|q_1\rangle
+ c_2\gamma|\psi_2(t)\rangle|q_2\rangle \nonumber\\
&& + (c_1\beta|\psi_1(t)\rangle
+ c_2\delta|\psi_2(t)\rangle)|q_3\rangle 
\label{fnstate}
\end{eqnarray}
where $|\psi_1(t)\rangle, |\psi_2(t)\rangle$ represent the wave-packets
appearing in (\ref{fstate}).
From (\ref{fnstate}) one can see that
the unambiguous path discrimination fails when one gets the state 
$|q_3\rangle$ while measuring the operator $\op{A}$. 
The probability of failure is just $|\langle q_3|\psi(t)\rangle|^2$
which turns out to be $c_1^2|\beta|^2+c_2^2|\delta|^2$, using the
orthogonality of $|\psi_1(t)\rangle, |\psi_2(t)\rangle$.
Subtracting that from 1, gives the {\em optimal} probability of 
unambiguously distinguishing between the two paths. Thus we can write
\begin{equation}
{\mathcal D}_Q = 1 - |\langle q_3|\psi(t)\rangle|^2,
\label{success1}
\end{equation}
which is our general expression for path distinguishability.
The distinguishability may also be calculated from the successful
discrimination as
\begin{equation}
{\mathcal D}_Q = |\langle q_1|\psi(t)\rangle|^2+|\langle q_2|\psi(t)\rangle|^2,
\label{success2}
\end{equation}
which would just be $c_1^2\alpha^2+c_2^2\gamma^2$.

\subsection{Case: $|\langle d_1|d_2\rangle|c_2/c_1 \ge |\langle d_1|d_2\rangle|^2$}

This is the case when the orthogonality of $|d_1\rangle, |d_2\rangle$ is
on the stronger side, and the asymmetry is not extreme. In this case the
values of $\alpha, \gamma$, for optimal success, are given by \cite{jaeger2}
\begin{eqnarray}
\alpha &=& \sqrt{1-|\langle d_1|d_2\rangle|c_2/c_1} \nonumber\\
\gamma &=& \sqrt{1-|\langle d_1|d_2\rangle|c_1/c_2}
\end{eqnarray}
Using (\ref{success2}), the distinguishability has the form
\begin{equation}
{\mathcal D}_Q = 1 - 2c_1c_2|\langle d_1|d_2\rangle|.
\label{DQ1}
\end{equation}
Using (\ref{DQ1}) and (\ref{V}) we arrive at the following 
relation
\begin{equation}
{\mathcal V} = \frac{2\sqrt{p_1p_2\xi}}{p_1+\xi p_2} \frac{(1-{\mathcal D}_Q)}
{2 c_1c_2}
\label{preduality1}
\end{equation}
Using (\ref{c1c2}), the above equation reduces to a very simple duality relation
\begin{equation}
{\mathcal D}_Q + {\mathcal V} = 1.
\label{duality1}
\end{equation}
This duality relation generalizes Englert's relation (\ref{englert}) to the case 
of asymmetric incoming quanton state, and is an equality, not an inequality for any pure state.
The relation (\ref{duality1}) implies that if one is able to unambiguously
distinguish between the two paths with a probability P {\em by any method},
that P cannot exceed ${\mathcal D}_Q$, and the fringe visibility cannot
exceed $1-{\mathcal D}_Q$.

If the state of the incoming quanton happens to be symmetric, i.e.,
$p_1=p_2=1/2$, and the two slits are of same width, i.e., $\xi=1$, we
can define a new distinguishability ${\mathcal D}$ as
\begin{equation}
{\mathcal D}^2 \equiv {\mathcal D}_Q(2-{\mathcal D}_Q)
= 1 - |\langle d_1|d_2\rangle|^2,
\end{equation}
which is precisely Englert's distinguishability \cite{englert}.
Using (\ref{V}) we can write
\begin{equation}
{\mathcal D}^2 + {\mathcal V}^2 = 1,
\end{equation}
which is just the saturated form of Englert's duality relation (\ref{englert}).
So for the symmetric case, (\ref{duality1}) is essentially the same as
(\ref{englert}).

\subsection{Case $|\langle d_1|d_2\rangle|^2 > |\langle d_1|d_2\rangle|c_2/c_1$}

This is the case when the orthogonality of $|d_1\rangle, |d_2\rangle$ is
on the lower side, and the asymmetry is large. In this case the values of
the constants are as follows \cite{jaeger2}
\begin{eqnarray}
\alpha &=& \sqrt{1-|\langle d_1|d_2\rangle|^2},~~~~
\beta=|\langle d_1|d_2\rangle| \nonumber\\
\gamma &=& 0,~~~~ |\delta| = 1.
\end{eqnarray}
The expression for distinguishability can be obtained by using (\ref{success2}):
\begin{equation}
{\mathcal D}_Q = c_1^2(1-|\langle d_1|d_2\rangle|^2).
\label{DQ2}
\end{equation}
Combining (\ref{DQ2}) and (\ref{V}), we can write
\begin{equation}
\frac{{\mathcal D}_Q}{c_1^2}
+ {\mathcal V}^2 \frac{(p_1+\xi p_2)^2}{4p_1p_2\xi} = 1,
\end{equation}
which can be rewriten as a new duality relation for this specific case:
\begin{equation}
\frac{{\mathcal D}_Q}{\tfrac{1}{2}(1+{\mathcal P}_0)} 
+ \frac{{\mathcal V}^2}{{\mathcal V}_0^2} = 1,
\label{duality2}
\end{equation}
where ${\mathcal V}_0$ is the {\em a priori} fringe visibility, and
${\mathcal P}_0$ is the {\em a priori} path-predictability defined as
${\mathcal P}_0 = \frac{|c_1|^2-|c_2|^2}{|c_1|^2+|c_2|^2}$ \cite{greenberger}.
As a consistency check, we consider the case where $|d_1\rangle,|d_2\rangle$
are identical, and hence ${\mathcal D}_Q$ given by (\ref{DQ2}) is zero.
Here the visibility is reduced to the {\em a priori} fringe visibility, 
as it should when there is no path-detection. Another special case is when
$p_1=1$, in which case ${\mathcal V}$ becomes zero, and (\ref{duality2})
gives ${\mathcal D}_Q=1$. Notice that varying the widths of the slits
affects the {\em a priori} fringe visibility and the {\em a priori}
path-predictability, but the equality (\ref{duality2}) continues to hold.

One might wonder if it is possible to have a single duality relation for
both the cases. To address this question, we denote the distinguishability
in the first case, i.e. (\ref{DQ1}), by $\mathcal{D}_{Q1}$ and that in
the second case (\ref{DQ2}), by $\mathcal{D}_{Q2}$. Then, in the region
$|\langle d_1|d_2\rangle|^2 > |\langle d_1|d_2\rangle|c_2/c_1$, one can show
that
\begin{equation}
\mathcal{D}_{Q1} - \mathcal{D}_{Q2} = c_1^2(|\langle d_1|d_2\rangle|
	- c_2/c_1)^2,
\end{equation}
which means that $\mathcal{D}_{Q2} \le \mathcal{D}_{Q1}$. This implies that
the following inequality holds in all regions
\begin{equation}
{\mathcal D}_Q + {\mathcal V} \le 1,
\end{equation}
but it cannot be saturated in the region 
$|\langle d_1|d_2\rangle|^2 > |\langle d_1|d_2\rangle|c_2/c_1$.

So we see that one cannot have a tight single duality relation for all
asymmetric two-slit experiments. Depending on the asymmetry and the
orthogonality of the path detector states, the duality relation has two
distinct forms, (\ref{duality1}) and (\ref{duality2}).

\subsection{The general case (pure/mixed)}

Till now we have been looking at the case where quanton and the path detector
are in a pure state. However, there are effects of decoherence due to which
there can be some loss of coherence, and it may become necessary to treat
the quanton and path detector combine as a mixed state. In such a situation,
the state of the quanton and path detector will be represented by a mixed
state density matrix. The treatment of path-distinguishability will remain
unchanged. For example, the path distinguishability given by (\ref{success1})
will now be represented as 
${\mathcal D}_Q =  1-\text{Trace}[\rho(t)|q_3\rangle\langle q_3|]$, and that
given by (\ref{success2}) will be represented as
${\mathcal D}_Q =  \text{Trace}[\rho(t)|q_1\rangle\langle q_1|]
+  \text{Trace}[\rho(t)|q_2\rangle\langle q_2|]$.

Fringe visibility is a measure of quantum coherence in the system, and
any mixedness will degrade the interference. This statement can be put
on a strong footing as follows. Recently a new measure of quantum coherence
was introduced, which, in a normalized form, can be written as
${\mathcal C} = {1\over n-1}\sum_{i\neq j} |\rho_{ij}|$. In our context,
$\rho_{ij}$ are the elements of the density matrix of the quanton,
$i,j$ corresponding to the two paths, and $n$ is the dimensionaility of the
Hilbert space (in our case $n=2$ corresponding to the two paths).
Using the final state of the quanton plus path detector as
$|\psi(t)\rangle = c_1|\psi_1(t)\rangle|d_1\rangle
+ c_2|\psi_2(t)\rangle|d_2\rangle$, we first trace over the path detector
states to get a reduced density matrix. The coherence ${\mathcal C}$
can then be evaluated, and turns out to be
\begin{equation}
{\mathcal C} = 2c_1c_2 |\langle d_1|d_2\rangle|.
\label{C}
\end{equation}
We see that for the two-slit interference, coherence is the same as visibility.
It has been shown that any incoherent operation on the system will lead to a
decrease in coherence ${\mathcal C}$ \cite{coherence}. In our case it means,
any mixedness introduced in the system will lead to a decrease in
the visibility ${\mathcal V}$.

Consequently, the visibility
will now be less than the maximum allowed by the amount of path information
that has been acquired by the path detector. For the case
$|\langle d_1|d_2\rangle| \le c_2/c_1$, it means
${\mathcal V} \le 1 - {\mathcal D}_Q $. Thus the duality relation becomes
the inequality
\begin{equation}
{\mathcal D}_Q + {\mathcal V} \le 1.
\label{gduality1}
\end{equation}
Similarly, for the case $|\langle d_1|d_2\rangle| > c_2/c_1$ too,
the duality relation becomes
\begin{equation}
\frac{{\mathcal D}_Q}{\tfrac{1}{2}(1+{\mathcal P}_0)} 
+ \frac{{\mathcal V}^2}{{\mathcal V}_0^2} \le 1.
\label{gduality2}
\end{equation}
The inequalities (\ref{gduality1}) and (\ref{gduality2}) quantify wave-particle 
duality for an asymmetric two slit interference. They are saturated for any
pure state.

\subsection{Particle or wave?}

The thought experiment in the preceding discussion, with an expanded Hilbert space,
was introduced to get an upper bound on the probability with which the
two paths can be unambiguously distinguished. However, if one were to actually
carry out this experiment with an observable $\op{A}$ of the path detector
giving three measured values, an interesting possibility emerges.
Suppose each quanton is detected on the screen in coincidence with measurement
of the observable $\op{A}$. Once the path detector is in place, the 
interference does not depend on what observable of the path detector we
choose to measure. Everytime we get the measured value 1, we know the quanton
went through slit 1, and everytime we get the value 2, we know that the
particular quanton went through slit 2. In these two situations, the quanton
behaves like a particle, choosing one of the two available paths. However,
when the measurment of $\op{A}$ yields the value 3, we conclude that the
quanton went through both the slits at the same time, behaving like a
spreadout wave. In fact, this can be experimentally verified by separating
the detected quantons into two groups, one where $\op{A}$ gave value 1 or 2,
and two where $\op{A}$ gave value 3. The first group of quantons will show
no interference, since path information for each of them is stored in the
path detector. The second group of quantons will show full interference.

The state of the quantons, for which measurement of $\op{A}$ gives value 3,
can be written using (\ref{fnstate}) as
\begin{eqnarray}
\langle q_3|\psi(t)\rangle &=& 
c_1\alpha|\psi_1(t)\rangle \langle q_3|q_1\rangle
+ c_2\gamma|\psi_2(t)\rangle \langle q_3|q_2\rangle \nonumber\\
&& + (c_1\beta|\psi_1(t)\rangle
+ c_2\delta|\psi_2(t)\rangle) \langle q_3|q_3\rangle \nonumber\\
&=&  c_1\beta|\psi_1(t)\rangle + c_2\delta|\psi_2(t)\rangle 
\label{wstate}
\end{eqnarray}
It is obvious that the above state will produce interference.
This leads us to conclude that in a two-slit experiment with an
{\em imperfect} path detector in place, each quanton can be thought of
as randomly choosing to behave like a particle or a wave. This behaviour,
obviously, is forced by the prescence of the path detector, in agreement
with the philosophy behind Bohr's principle of complementarity \cite{bohr}.

\section{Conclusion}

In conclusion, we have analyzed the issue of wave-particle duality in a
two-slit experiment. For symmetric beams and equal slit widths, wave-particle
duality can be captured by the well-known inequality (\ref{englert}), which
was derived using the ideas of minimum error discrimination of states
\cite{englert}. The same relation can be derived by defining the
distinguishability using the ideas of UQSD \cite{3slit}. This latter
method has proved to be very useful in describing wave-particle duality
in multi-slit interference \cite{3slit,cd,nslit}. For two-slit experiments
where the two beams are asymmetric, and the slits may be of unequal widths,
a result as strong as (\ref{englert}) was lacking. We have used this new
approach to study wave-particle duality in this asymmetric case.

We argue that in a two-slit experiment, getting path
information should mean, being able to tell unambiguously  for each quanton,
which of the two slits it went through. Using this premise, we use a 
thought experiment to get which path information about the quantons using
UQSD. We define the path distinguishability as the maximum probability with
which one can unambguously tell which slit the quanton went through,
{\em in principle}. Using it we derive two duality relations for intrference
where the two paths may not be equally probable or the two slits may not
be of equal widths. The two duality relations
correspond to two difference ranges of asymmetry. Unlike the well studied
symmetric case, a single tight duality relation is not possible for the
asymmetric case.  Additionally, if the
thought experiment is actually performed, one can tell for each quanton
if it went through slit 1 or slit 2 like a particle or through both the slits
like a wave.

\section*{Acknowledgement}
Keerthy Menon is thankful to the Centre for Theoretical Physics, Jamia Millia
Islamia for providing the facilities of the Centre during the course of 
this work.

\end{document}